\documentclass{article}

\usepackage{arxiv}

\usepackage[utf8]{inputenc} % allow utf-8 input
\usepackage[T1]{fontenc}    % use 8-bit T1 fonts
\usepackage{hyperref}       % hyperlinks
\usepackage{url}            % simple URL typesetting
\usepackage{booktabs}       % professional-quality tables
\usepackage{amsfonts}       % blackboard math symbols
\usepackage{nicefrac}       % compact symbols for 1/2, etc.
\usepackage{microtype}      % microtypography
\usepackage{lipsum}		% Can be removed after putting your text content
\usepackage{setspace}
\usepackage{graphicx}
\usepackage{doi}
\usepackage{amsmath}
\usepackage[numbers,sort&compress]{natbib}

\title{Exploring the Potential of Two-dimensional Borospherene for Toxic Gas Sensing and Capture: A DFT Study}

\author{%
\parbox{0.95\linewidth}{\centering
\href{https://orcid.org/0000-0001-7653-0428}{\includegraphics[scale=0.09]{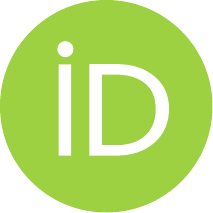}\hspace{1mm}}Nicolas F.~Martins\textsuperscript{1},
\href{https://orcid.org/0000-0002-8366-7227}{\includegraphics[scale=0.09]{icons/orcid.pdf}\hspace{1mm}}Jos\'e A.~dos S.~Laranjeira\textsuperscript{1},
\href{https://orcid.org/0000-0003-4699-5886}{\includegraphics[scale=0.09]{icons/orcid.pdf}\hspace{1mm}}Kleuton A.~L.~Lima\textsuperscript{2},
\href{https://orcid.org/0000-0001-7468-2946}{\includegraphics[scale=0.09]{icons/orcid.pdf}\hspace{1mm}}Luiz A.~Ribeiro Jr\textsuperscript{3,$\dag$},
and
\href{https://orcid.org/0000-0002-5217-7145}{\includegraphics[scale=0.09]{icons/orcid.pdf}\hspace{1mm}}Julio R. Sambrano\textsuperscript{1},
 \\
\vspace{0.6em}
{\normalfont\normalsize
\textsuperscript{1}Modeling and Molecular Simulation Group, S\~ao Paulo State University (UNESP), School of Sciences, Bauru 17033-360, SP, Brazil\\
\textsuperscript{2}Department of Applied Physics and Center for Computational Engineering and Sciences, State University of Campinas, Campinas, 13083-859, SP, Brazil\\
\textsuperscript{3}Computational Materials Laboratory, LCCMat, Institute of Physics, University of Bras\'ilia, 70910-900, Bras\'ilia, Federal District, Brazil\\
\vspace{0.6em}
\href{https://scholar.google.com/citations?user=EBMMpbYAAAAJ&hl=pt-BR}{\includegraphics[scale=0.05]{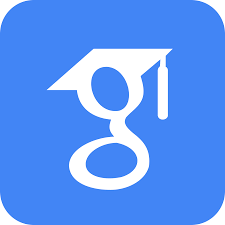}} \href{https://www.linkedin.com/in/nicolas-martins-765136388/}{\includegraphics[scale=0.05]{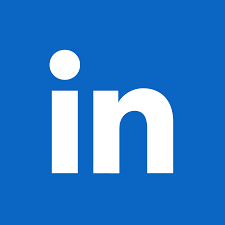}}\hspace{0.1cm}\texttt{\textsuperscript{*}nicolas.ferreira@unesp.br } \\
\vspace{0.1cm}
\href{https://scholar.google.com/citations?user=Gs2UkTgAAAAJ&hl=pt-BR}{\includegraphics[scale=0.05]{icons/gscholar.png}} \href{https://www.linkedin.com/in/julio-sambrano-0257591a5/}{\includegraphics[scale=0.05]{icons/linkedin.png}}\hspace{0.1cm}\texttt{\textsuperscript{$\dag$}jr.sambrano@unesp.br }\\
}
}%
}
\renewcommand{\shorttitle}{Nicolas F.~Martins et al.}

\hypersetup{
pdftitle={3d-dodeca},
pdfsubject={3d-dodeca},
pdfauthor={Kleuton A. L. Lima, Luiz A. Ribeiro Jr},
pdfkeywords={Two-dimensional metals, Kagome lattice, Goldene, Lattice dynamics, Strain engineering, First-principles calculations.},
}

\begin{document}
\maketitle

\onehalfspacing

\begin{abstract}
Two-dimensional (2D) boron-based materials have gained increasing interest due to their exceptional physicochemical properties and potential technological applications.  In this way, borospherenes, a 2D  Boron-based fullerene-like lattice (2D–B\textsubscript{40}), are explored due to their potential for capturing and detecting toxic gases, such as CO, NO, NH3, and SO2. Therefore, density functional theory simulations were carried out to explore the adsorption energy and the distinct interaction regimes, where CO exhibits weak physisorption ($-0.16$ eV), while NO ($-2.24$ eV), NH$_3$ ($-1.47$ eV), and SO$_2$ ($-1.51$ eV) undergo strong chemisorption. Bader charge analysis reveals significant electron donation from 2D–B\textsubscript{40} to NO and electron acceptance from SO$_2$. These interactions cause measurable shifts in work function, with SO$_2$ producing the most significant modulation ($\Delta\Phi = +14.6\%$). Remarkably, \textit{ab initio} molecular dynamics simulations (AIMD) reveal spontaneous SO$_2$ decomposition at room temperature, indicating dual functionality for both sensing and environmental remediation. Compared to other boron-based materials, such as $\chi_3$-borophene, $\beta_{12}$-borophene, and B$_{40}$ fullerene, 2D–B\textsubscript{40} exhibits superior gas affinity, positioning it as a versatile platform for the detection and capture of toxic gases.
\end{abstract}

% keywords can be removed

\section{Introduction}

Two-dimensional (2D)  materials have garnered increasing attention in recent years due to their physicochemical properties and potential technological applications in nanoelectronics, energy storage, and gas sensing~\cite{zhang2017two,wu2012two, huzaifa2025exploring, martins2025exploring, martins2025hop, laranjeira2025hydrogen}. Among these, borophene and its various polymorphs have been widely explored as platforms for gas detection and capture~\cite{hou2022borophene,shukla2017toward,mohanty2023review}, owing to their high surface area, structural anisotropy, and tunable electronic structure~\cite{ou2021emergence,hou2020borophene,sun2017two}. Notably, the $\chi$\textsubscript{3} and $\beta$\textsubscript{12} phases have been extensively studied~\cite{zergani2022gas,sun2022designing}, showing differentiated interactions with adsorbates through mechanisms such as orbital hybridization and charge transfer.

Several theoretical studies based on density functional theory (DFT) have demonstrated the remarkable gas adsorption properties of borophene derivatives. For instance, NO and SO\textsubscript{2} molecules are known to chemisorb on $\chi$\textsubscript{3} strongly and $\beta$\textsubscript{12}-borophene, while gases such as CO and NH\textsubscript{3} exhibit weaker interactions dominated by van der Waals forces~\cite{liu2017first,kumar2020enhancing,zergani2022gas,ta2021adsorption,sabokdast2023prediction}. In particular, the functionalization of borophene with alkali metals has been shown to enhance gas capture via increased charge transfer and work function modulation~\cite{tu2021first}.

Liu \textit{et al.}~\cite{liu2017first} and Zergani \textit{et al.}~\cite{zergani2022gas} reported that NO and SO\textsubscript{2} molecules exhibit strong chemisorption on $\chi$\textsubscript{3}- and $\beta$\textsubscript{12}-borophene, while CO and NH\textsubscript{3} display weaker physisorption, primarily governed by van der Waals interactions. The construction of borophene-based heterostructures with MoS\textsubscript{2}~\cite{mohanty2023review} has shown improved gas adsorption capacity and structural robustness. On the other hand, Wang \textit{et al.}~\cite{WANG2022111033} demonstrated that Li-doping enhances sulfur gas adsorption on borophene–graphene composites while maintaining structural integrity. Likewise, Arefi \textit{et al.}~\cite{AREFI2021113159} found that Na-decorated borophene significantly increases the adsorption energy of CO\textsubscript{2}, converting a physisorptive interaction into a chemisorptive one, with concurrent changes in current–voltage characteristics. In another approach, Arkoti and Pal~\cite{9967146} showed that Ba-decoration modulates the work function of borophene, enhancing its sensitivity to NO\textsubscript{2} at room temperature. Anversa \textit{et al.}~\cite{ANVERSA2023122370} explored Pt-decorated borophene, where hazardous gases such as NO and SO\textsubscript{2} strongly interact with the Pt site, altering the electronic states near the Fermi level and suggesting selectivity based on charge transfer and work function variation.

These results have motivated the exploration of alternative boron-based 2D architectures~\cite{https://doi.org/10.1002/cctc.202301527, Kumar2024}. A notable example is the recently proposed 2D-B\textsubscript{40} monolayer, a theoretically predicted planar extension of the boron fullerene B\textsubscript{40} cage~\cite{zhai2014observation,mortazavi2023first}. This network consists of a periodic tiling of B\textsubscript{40}-like units, forming a 2D lattice with a unit cell length of 7.64~\AA{} and a cohesive energy of $-6.23$~eV/atom, which is $0.15$~eV/atom lower than the isolated B\textsubscript{40} cage, indicating favorable energetic stability~\cite{mortazavi2023first}. Mechanical and thermal analyzes showed an elastic modulus of 125~N/m, a tensile strength of 7.8~N/m, and structural stability maintained up to 700~K in \textit{ab initio} molecular dynamics (AIMD) simulations. The Poisson's ratio is remarkably low ($\nu = 0.010$), suggesting minimal transverse deformation under uniaxial strain. 

In this work, the first comprehensive DFT investigation of the adsorption behavior of toxic gas molecules, such as CO, NO, NH\textsubscript{3}, and SO\textsubscript{2}, on the 2D-B\textsubscript {40} monolayer is presented. Charge density difference (CDD) mapping and Bader analysis were employed to analyze the charge redistribution. Changes in the electronic structure and work function were also evaluated to assess the suitability of 2D–B\textsubscript{40} as a sensor material. Furthermore, AIMD simulations at 300~K were performed to verify the structural resilience of the system under gas exposure. The findings highlight the robust adsorption performance and chemical reactivity of 2D–B\textsubscript{40}, positioning it as a candidate for next-generation gas sensing and environmental remediation technologies.

\section{Computational Setup}

DFT simulations were performed using the Vienna Ab Initio Simulation Package (VASP) \cite{kresse1999ultrasoft}. Projected augmented wave (PAW) potentials and the Perdew-Burke-Ernzerhof (PBE) \cite{PhysRevLett.77.3865} generalized gradient approximation (GGA) were employed \cite{PhysRevB.50.17953}. A kinetic energy cutoff of 520 eV was used to ensure accuracy in the calculations. During structural optimization, the K-mesh was set to 5 $\times$ 5 $\times$ 1, while density of states (DOS) calculations were performed with a $\Gamma$-centered K-point grid of 8 $\times$ 8 $\times$ 1. To account for long-range interactions, Grimme's DFT-D3 \cite{grimme2006semiempirical} dispersion correction was included. Structural optimization was performed using the conjugate gradient algorithm until the convergence criteria were satisfied. 

The energy errors for atomic positions and lattice parameters were required to be less than $1 \times 10^{-5}$ eV, and the Hellmann-Feynman forces on each atom had to be within 0.01 eV/\r{A}. A 20 \r{A} vacuum along the z-axis was applied to avoid interactions between periodic images in the transverse direction. To evaluate the dynamical stability of the 2D B\textsubscript{40} fullerene network, phonon dispersion calculations were performed using density functional perturbation theory (DFPT) as implemented in the Phonopy package \cite{togo2015first}. Additionally, AIMD simulations were conducted at 300 K for 5 ps in the NVT ensemble, employing the Nos\'e-Hoover thermostat \cite{hoover1985canonical} to assess its thermal robustness.

\section{Results and Discussion}

\subsection{Structural and electronic description of Borospherene (2D–B\textsubscript{40})}

The 2D-B\textsubscript{40} network has a tetragonal unit cell with lattice constants \(a = b = 7.60\) \r{A}, in agreement with the value reported by Mortazavi \cite{mortazavi2023first}. Within this framework, each boron–boron bond measures, on average, 1.72 \r{A}, slightly longer than in planar borophene (1.67 \r{A})\cite{yang2008ab} and comparable to the B\textsubscript{40} fullerene cage (1.74 \r{A})\cite{yang2019aggregation}, reflecting the mixture of hexagonal and heptagonal motifs that stabilize the sheet. 

\begin{figure}[!htb]
    \centering
    \includegraphics[width=\linewidth]{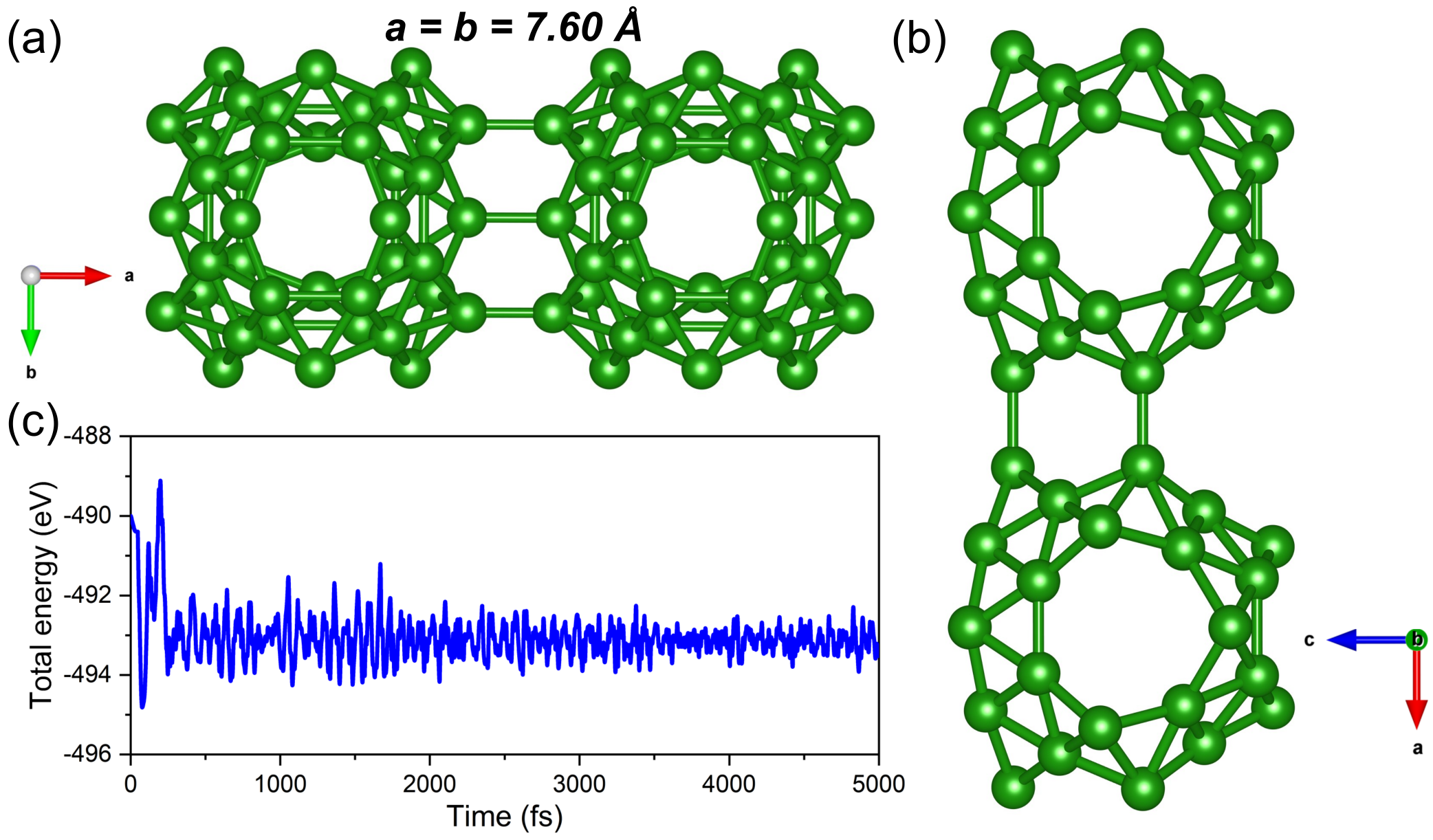}
     \caption{Structural and thermal stability analysis of the 2D–B\textsubscript{40} network. (a) Top view and (b) side view of a \(2\times2\times1\) supercell illustrating the alternation of hexagonal and heptagonal boron rings (green spheres). (c) Total energy evolution during 5 ps of AIMD simulation run at 300 K.}
    \label{fig:1}
\end{figure}

Figure~\ref{fig:1}a–b shows the top and side views of a \(2\times2\times1\) supercell of 2D–B\textsubscript{40}, where the periodic arrangement of 6- and 7-membered rings is present.  AIMD simulations were utilized to inspect the thermal stability of the material. (Fig.~\ref{fig:1}c). The total energy oscillates within \(\pm2\) eV of its mean value (–493 eV), with no drift or bond breaking observed, confirming that the 2D-B\textsubscript{40} network remains intact under ambient thermal conditions. 

\begin{figure}[!htb]
    \centering
    \includegraphics[width=\linewidth]{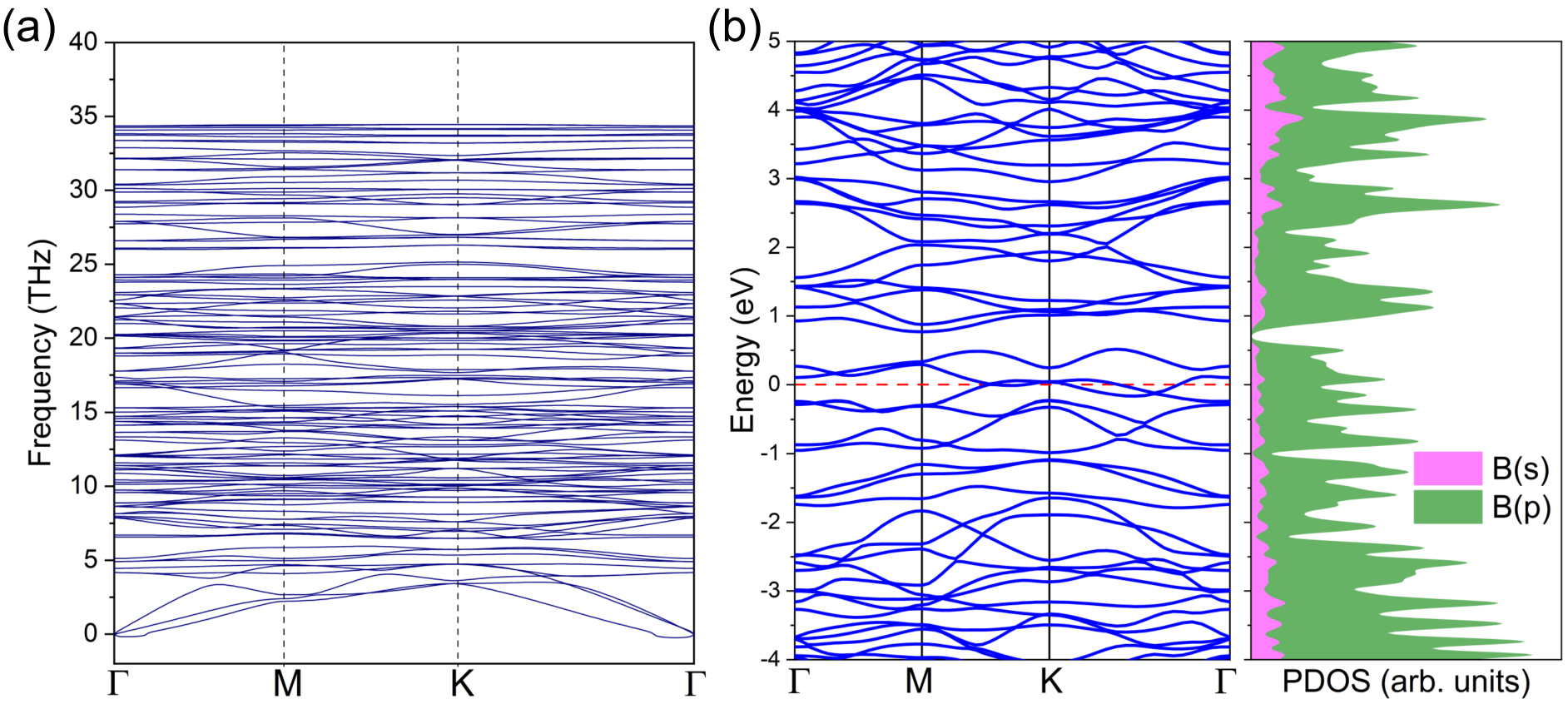}
     \caption{(a) Phonon dispersion of the 2D–B\textsubscript{40} network along $\Gamma\text{–}M\text{–}K\text{–}\Gamma$, showing the absence of imaginary modes and confirming dynamical stability. (b) Electronic band structure (left) plotted with the Fermi level at 0 eV (red dashed line) and the projected density of states (right).}
    \label{fig:2}
\end{figure}

 As shown in Figure~\ref{fig:2}(a), the phonon dispersion along the high‐symmetry path $\Gamma\text{–}M\text{–}K\text{–}\Gamma$ exhibits no imaginary modes, confirming that the B\textsubscript{40} lattice is dynamically stable. The highest optical branches reach approximately 38 THz, consistent with well-reported boron-based 2D structures~\cite{10.1063/1.4953775, Xiao2017}.

Figure~\ref{fig:2}(b) depicts the electronic band structure with the Fermi level set to 0 eV (red dashed line). Two bands cross the Fermi level, indicative of the metallic character of the network and the high density of conduction channels. To the right, the projected density of states (PDOS) decomposes the spectral weight into B($s$) and B($p$) contributions. The PDOS shows that states near the Fermi level are dominated by B($p$) orbitals. In contrast, B($s$) states are distributed across the entire energy range with minor contributions, underscoring the role of $p$–orbital hybridization in mediating the conductivity.

\subsection{Toxic gas adsorption}

\begin{figure}[htb!]
    \centering
    \includegraphics[width=0.75\linewidth]{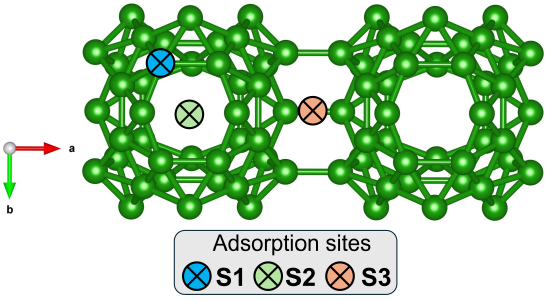}
     \caption{Top view of the 2D–B\textsubscript{40} network highlighting the three adsorption sites considered: S1 (blue cross) atop a boron atom, S2 (green cross) at the center of a heptagonal ring, and S3 (orange cross) at a bridge between two boron atoms. The crystallographic axes $a$ and $b$ are indicated.}
    \label{fig:sites}
\end{figure}

To evaluate the gas adsorption of (CO, NO, NH$_3$, and SO$_2$), three representative adsorption sites (S1, S2, and S3) were selected, as shown in Figure~\ref{fig:sites}. The adsorption energy ($E_\mathrm{ads}$) was computed to assess the relative stability of each configuration, using the following equation: 
\begin{equation}
E_\mathrm{ads} = E_{\text{gas+B}_{40}} - E_{\text{B}_{40}} - E_{\text{gas}},
\end{equation}
\noindent where $E_{\text{gas+B}_{40}}$ is the total energy of the relaxed gas–B\textsubscript{40} system, $E_{\text{B}_{40}}$ is the energy of the isolated monolayer, and $E_{\text{gas}}$ is the energy of the isolated gas molecule \textbf{in vacuum.} 

\begin{figure} [htb!]
    \centering
    \includegraphics[width=0.75\linewidth]{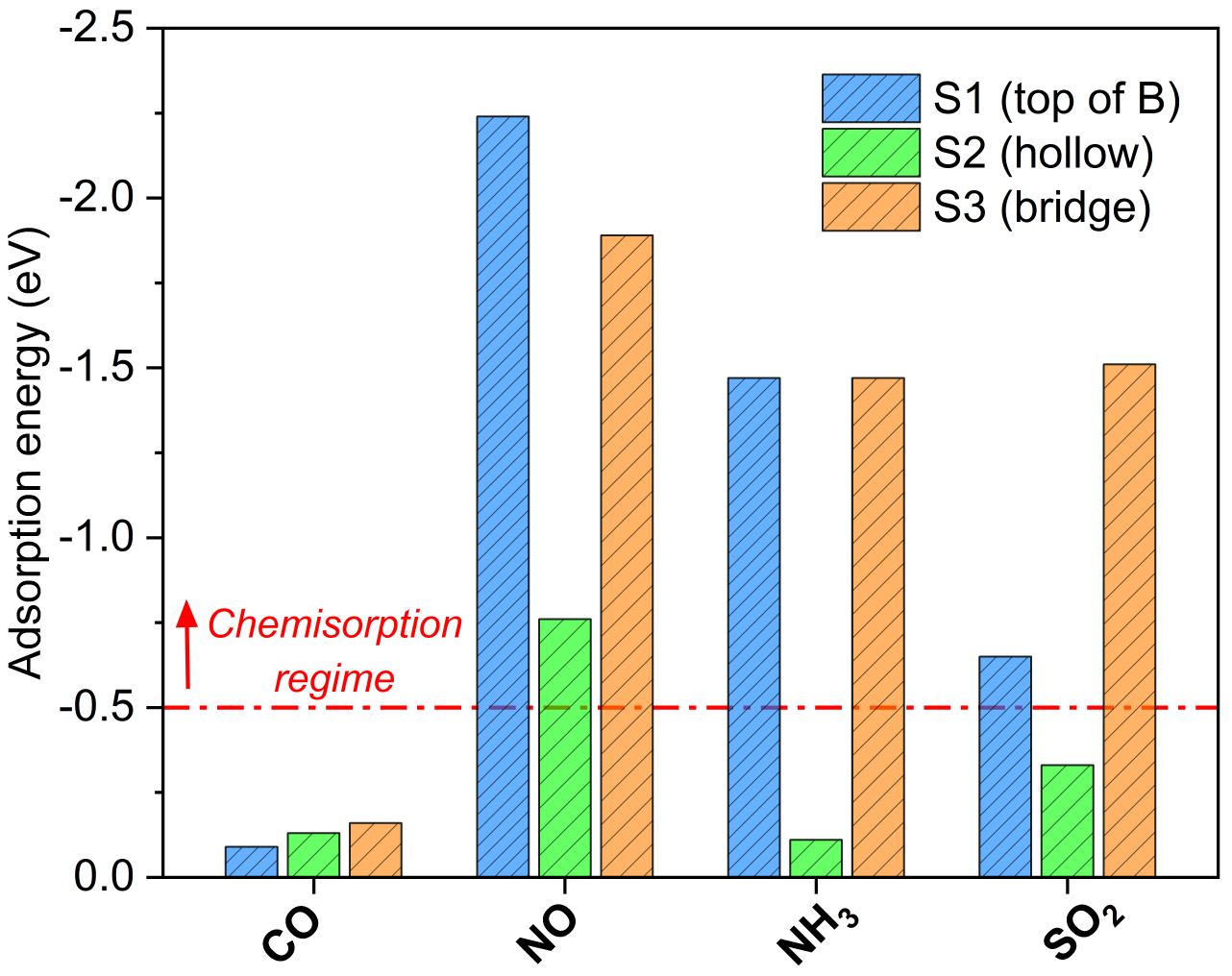}
     \caption{Adsorption energies (\(E_\mathrm{ads}\)) of CO, NO, NH\(_3\) and SO\(_2\) at sites S1 (top of B), S2 (hollow center) and S3 (bridge) on the 2D–B\(_{40}\) network. The red dashed line at \(|E_\mathrm{ads}|=0.5\) eV separates physisorption (above) from chemisorption (below).}
    \label{fig:adsorption result}
\end{figure}

Figure~\ref{fig:adsorption result} summarizes the $E_\mathrm{ads}$ values calculated for CO, NO, NH\textsubscript{3}, and SO\textsubscript{2} on the selected sites. Among them, CO exhibits weak binding, with a minimum energy of $-0.16$~eV at the bridge site (S3), and a relatively large equilibrium separation of 2.96~\r{A}, suggesting a physisorptive interaction. In contrast, NO, NH\textsubscript{3}, and SO\textsubscript{2} show significantly larger binding energies, all below $-1.4$~eV, consistent with stronger interactions. NO reaches its minimum energy ($-2.24$~eV) at the on-top site S1, whereas NH\textsubscript{3} and SO\textsubscript{2} favor the bridge site (S3), with adsorption energies of $-1.47$~eV and $-1.51$~eV, respectively. Here, we consider that an $E_\mathrm{ads}$ with an absolute value greater than 0.5~eV indicates a chemisorption regime, whereas lower values correspond to physisorption~\cite{gergen2001chemically}. However, it is essential to investigate the charge transfer mechanism to validate this observation, which will be discussed in the following analyses.

\begin{figure}[htb!]
    \centering
    \includegraphics[width=1\linewidth]{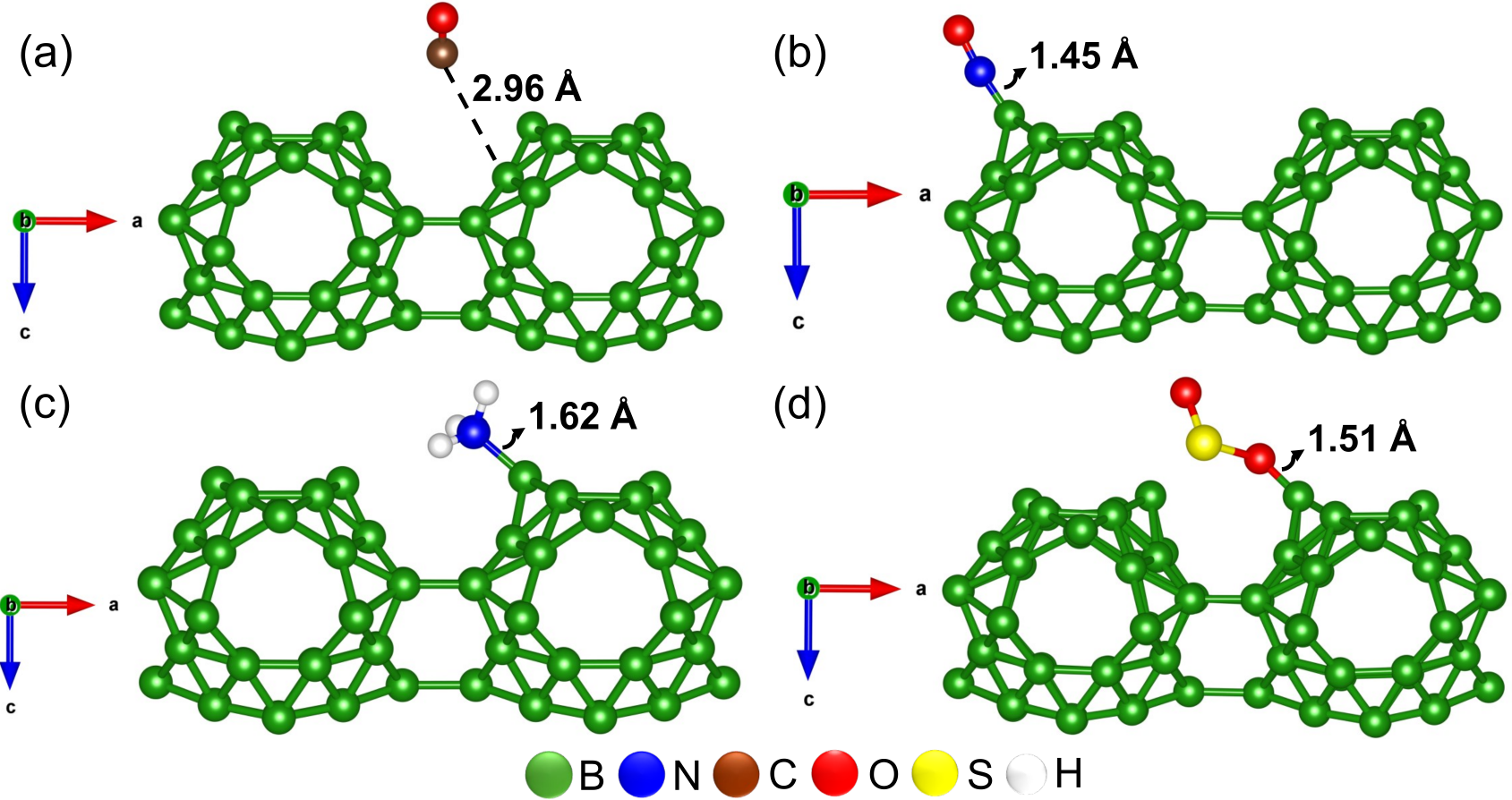}
    \caption{Side views of the most stable adsorption geometries on the 2D–B\textsubscript{40} network: (a) CO at the S3 bridge site (B–C distance 2.96 \r{A}); (b) NO atop a boron atom at S1 (B–N distance 1.45 \r{A}); (c) NH\(_3\) at S3 (B–N distance 1.62 \r{A}); and (d) SO\(_2\) at S3 (B–S distance 1.51 \r{A}).}
    \label{fig:ads_geoms}
\end{figure}

The optimized adsorption geometries in Figure~\ref{fig:ads_geoms} show the atomic binding configurations that support the adsorption energies discussed previously. In panel (a), CO resides above the bridge site (S3) at a distance of 2.96 \r{A}, consistent with its weak physisorption energy of –0.16 eV. Panel (b) shows NO chemisorbed atop a boron atom (S1) with a B–N bond length of 1.45 \r{A}, in agreement with its strong adsorption energy of –2.24 eV. Panels (c) and (d) depict NH\(_3\) and SO\(_2\) at the bridge site (S3), where the B–N distance of 1.62 \r{A} and B–S distance of 1.51 \r{A}, respectively, reflect their moderate chemisorption energies (–1.47 eV and –1.51 eV). Note that in each case the molecule’s center of mass shifts slightly toward the nearest boron atom upon relaxation, highlighting the directional character of the chemisorptive bonds on the 2D-B\(_{40}\) network.

\begin{table}[h!]
\centering
\setlength{\tabcolsep}{12pt} % aumenta o espaço entre colunas
\renewcommand{\arraystretch}{1.3} % aumenta a altura das linhas
\begin{tabular}{lcccc}
\hline
\textbf{Boron-based material} & \multicolumn{4}{c}{\textbf{E$_\mathrm{ads}$ (eV)}} \\
\cline{2-5}
 & CO & NO & NH$_3$ & SO$_2$ \\
\hline
2D-B\(_{40}\) (this work) & -0.16 & -2.24 & -1.47 & -1.51 \\
Stripped Borophene\cite{liu2017first, tu2021first}  & -0.76 & -4.04 & -1.96 & -1.36 \\
$\chi^3$-Borophene\cite{duan2023first, zergani2022gas} & -0.84 & -1.44 & -0.76 & -1.37 \\
$\beta^{12}$-Borophene\cite{huang2018adsorption} & -1.19 & -0.95 & -1.11 & - \\
B$_{35}$ Nanocluster\cite{hossain2020ab} & - & -0.33 & -0.18 & - \\
B$_{40}$ Fullerene\cite{lin2016b40} & - & - & -1.09 & - \\
\hline
\end{tabular}
\caption{Adsorption energies (E$_\mathrm{ads}$) of various boron-based materials with different molecules.}
\label{tab:boron_adsorption}
\end{table}

Table~\ref{tab:boron_adsorption} summarizes the adsorption energies on 2D–B\textsubscript{40} alongside values reported for other boron‐based materials, including pristine borophene, the $\chi$\textsubscript{3} and $\beta$\textsubscript{12} polymorphs, a B\textsubscript{35} nanocluster, and the B\textsubscript{40} fullerene. Liu \emph{et al.} found that pristine borophene chemisorbs CO, NO, and NH\(_3\) with \(E_{\mathrm{ads}}=-0.76\), \(-4.04\) and \(-1.96\)\,eV, respectively, while SO\(_2\) adsorption reaches \(-1.36\)\,eV on Li–decorated borophene \cite{liu2017first,tu2021first}. DFT studies on $\chi$\textsubscript{3}-borophene report adsorption energies of \(-1.44\)\,eV for NO and \(-0.84\)\,eV for CO \cite{zergani2022gas,duan2023first}, and $\beta$\textsubscript{12}-borophene shows similar chemisorption strengths for NH\(_3\) (–1.11 eV) and SO\(_2\) (–1.36 eV) \cite{huang2018adsorption}. In comparison, our 2D–B\textsubscript{40} network binds NO, NH\(_3\) and SO\(_2\) more strongly (–2.24, –1.47 and –1.51 eV, respectively), and even surpasses the binding affinities reported for B\textsubscript{35} clusters (–0.98 to –1.20 eV) \cite{hossain2020ab} and the B\textsubscript{40} fullerene (–1.20 to –1.35 eV) \cite{lin2016b40}. Additionally, the results unveiled here are comparable, in terms of binding strength, with adsorption values for other 2D platforms \cite{MARTINS2025109777, shi2023adsorption, liu2023adsorption, mistry2025adsorption}. These comparisons highlight the superior gas-capture performance of the 2D borospherene network.  

\begin{figure}[htb!]
    \centering
    \includegraphics[width=0.75\linewidth]{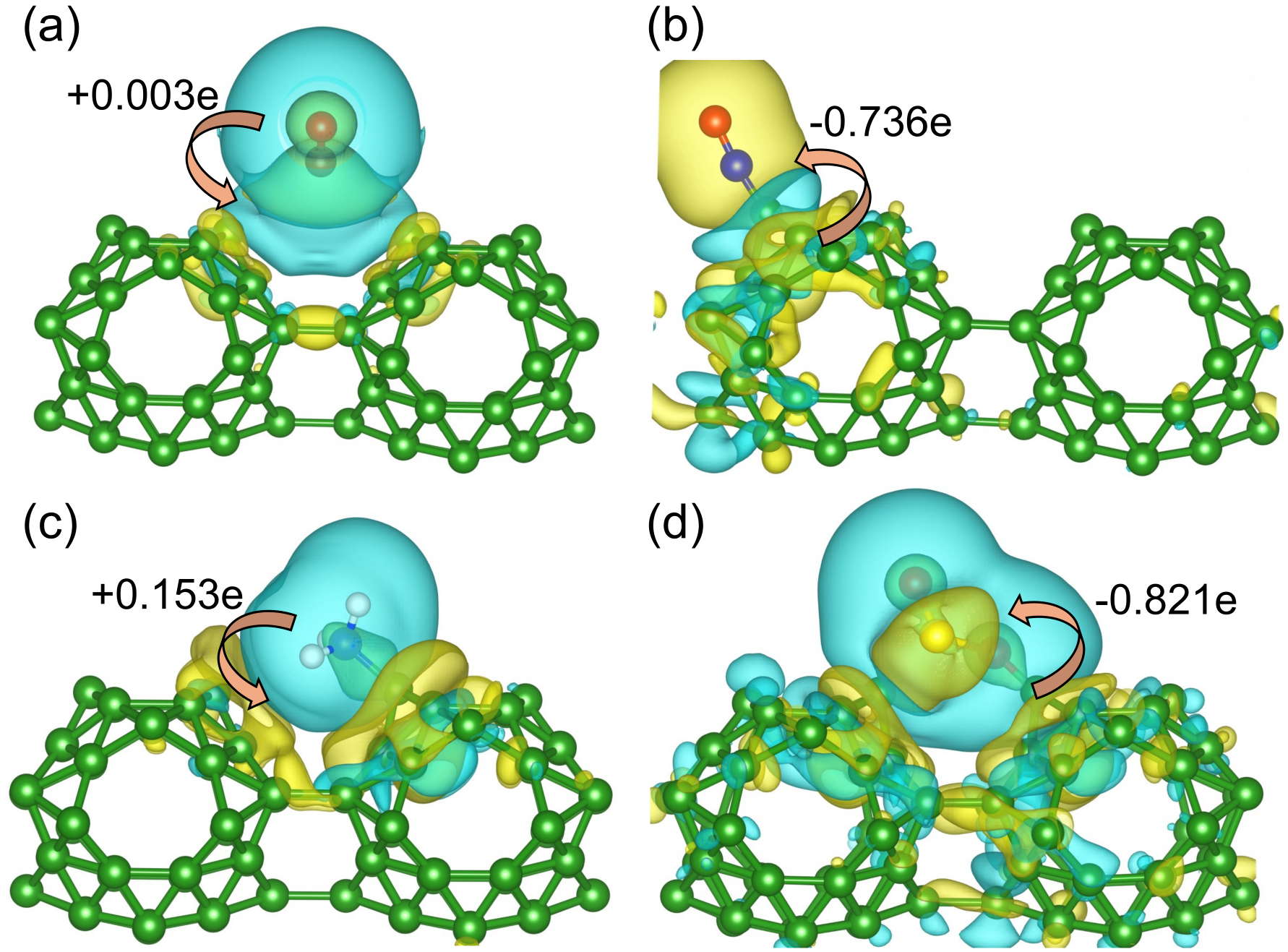}
    \caption{Charge‐density difference (CDD) isosurfaces for adsorption on 2D–B\textsubscript{40}, plotted at \(\pm0.002\)\,\(e/\text{\r{A}}^3\). Cyan regions denote depletion, and yellow regions denote accumulation of electronic charge. Net Bader charge transfer \(\Delta Q\) is indicated for (a) CO (+0.003 \(e\)), (b) NO (–0.736 \(e\)), (c) NH\(_3\) (+0.153 \(e\)), and (d) SO\(_2\) (–0.821 \(e\)).}
   \label{fig:CDD}
\end{figure}

To complement the adsorption energy analysis, Bader charge analysis and charge density difference (CDD) maps were computed for CO, NO, NH$_3$, and SO$_2$ adsorbed on the 2D–B\textsubscript{40} monolayer. The CDD was obtained using the following expression:  \begin{equation} \Delta \rho(\mathbf{r}) = \rho_{\text{gas+B}_{40}}(\mathbf{r}) - \rho_{\text{B}_{40}}(\mathbf{r}) - \rho_{\text{gas}}(\mathbf{r}),
\end{equation}

\noindent where $\rho_{\text{gas+B}_{40}}$ is the charge density of the relaxed adsorbate–substrate system, $\rho_{\text{B}_{40}}$ is the charge density of the clean B\textsubscript{40} surface, and $\rho_{\text{gas}}$ is the charge density of the isolated gas molecule. Figure 6 shows ~\ref{fig:CDD} shows the relaxed geometries overlaid with isosurfaces of charge depletion (cyan) and accumulation (yellow), as well as the net charge transfer \(\Delta Q\) for each adsorbate. CO (panel a) transfers only +0.003 \(e\) to the substrate, consistent with its weak physisorption. NO (panel b) accepts -0.736 \(e\), forming pronounced depletion lobes on adjacent boron atoms and indicating strong chemisorption. NH$_3$ (panel c) donates +0.153 \(e\), evidenced by localized charge accumulation around the boron centers. SO$_2$ (panel d) withdraws -0.821 \(e\), producing extensive depletion regions at the S–B interface. These distinct charge-transfer signatures corroborate the selective chemisorptive behavior of NO and SO$_2$ versus the physisorptive interaction of CO. 

\begin{figure} [ht!]
    \centering
    \includegraphics[width=0.75\linewidth]{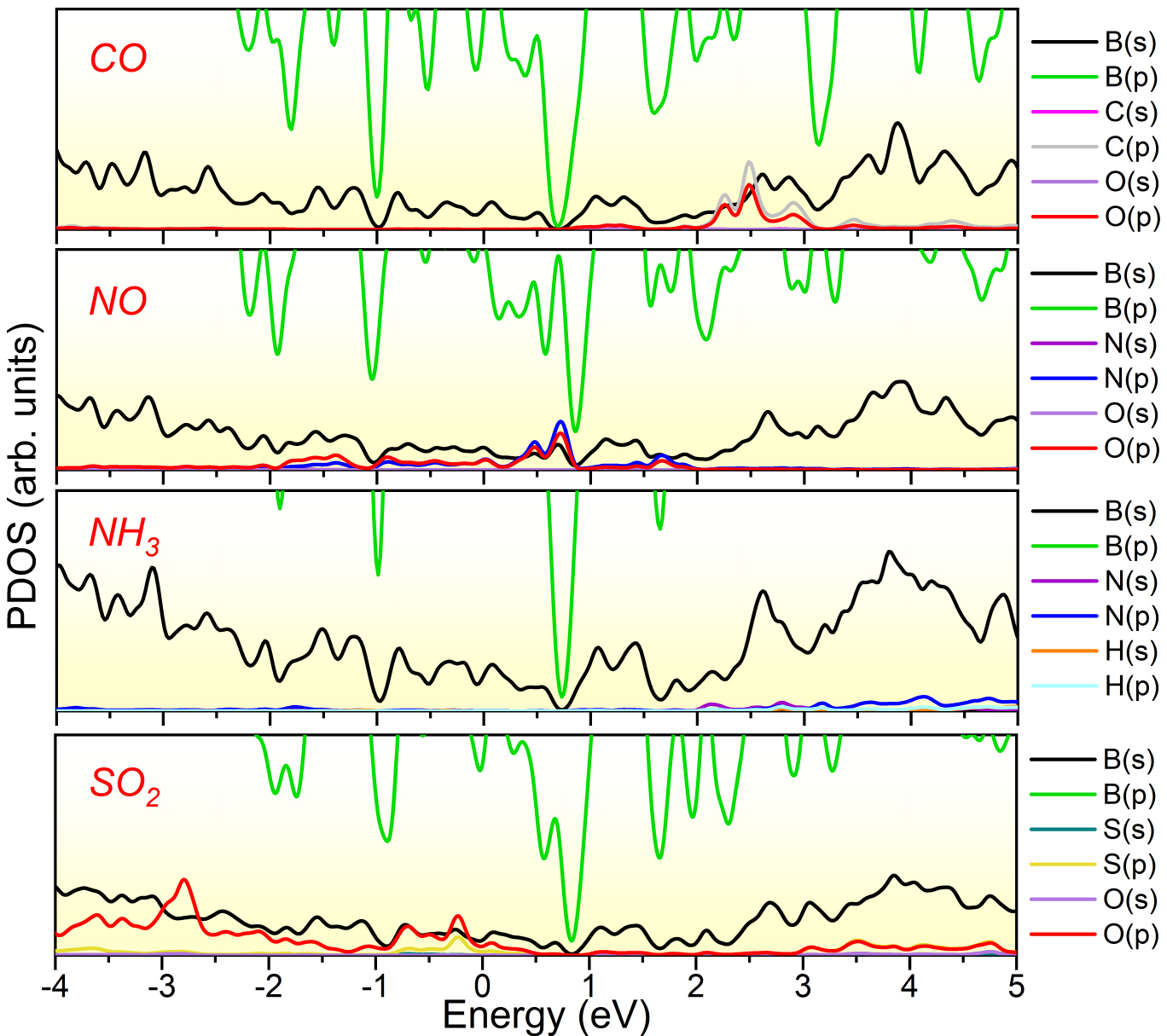}
     \caption{Projected density of states (PDOS) for CO, NO, NH\(_3\) and SO\(_2\) adsorbed on 2D–B\textsubscript{40}, with the Fermi level at 0 eV.}
  \label{fig:PDOS gases}
\end{figure}

The projected density of states (PDOS), presented in Figure~\ref{fig:PDOS gases}, highlights the orbital interactions that underpin gas binding on 2D–B\textsubscript{40}. For CO adsorption, the C($p$) and O($p$) states appear near the Fermi level but exhibit only weak overlap with B($s$) and B($p$) bands, consistent with a van der Waals–dominated physisorption regime. In the NO–boron system, strong B($p$)–N($p$) hybridization emerges between 0 and 1 eV above the Fermi level, and additional mixing in the –2 to –0.5 eV range, corroborating the significant chemisorption energy and the formation of covalent B–N bonds. NH\(_3\) adsorption shows modest mixing of N($p$) with B($p$) below the Fermi level, reflecting its intermediate binding strength and partial charge donation. Finally, SO\(_2\) exhibits pronounced S($p$) and O($p$) contributions spanning -3 to 0 eV, with extensive overlap with B($p$) states, underscoring the strong chemisorptive character and significant charge transfer in this system.

\begin{figure} [htb!]
    \centering
    \includegraphics[width=0.5\linewidth]{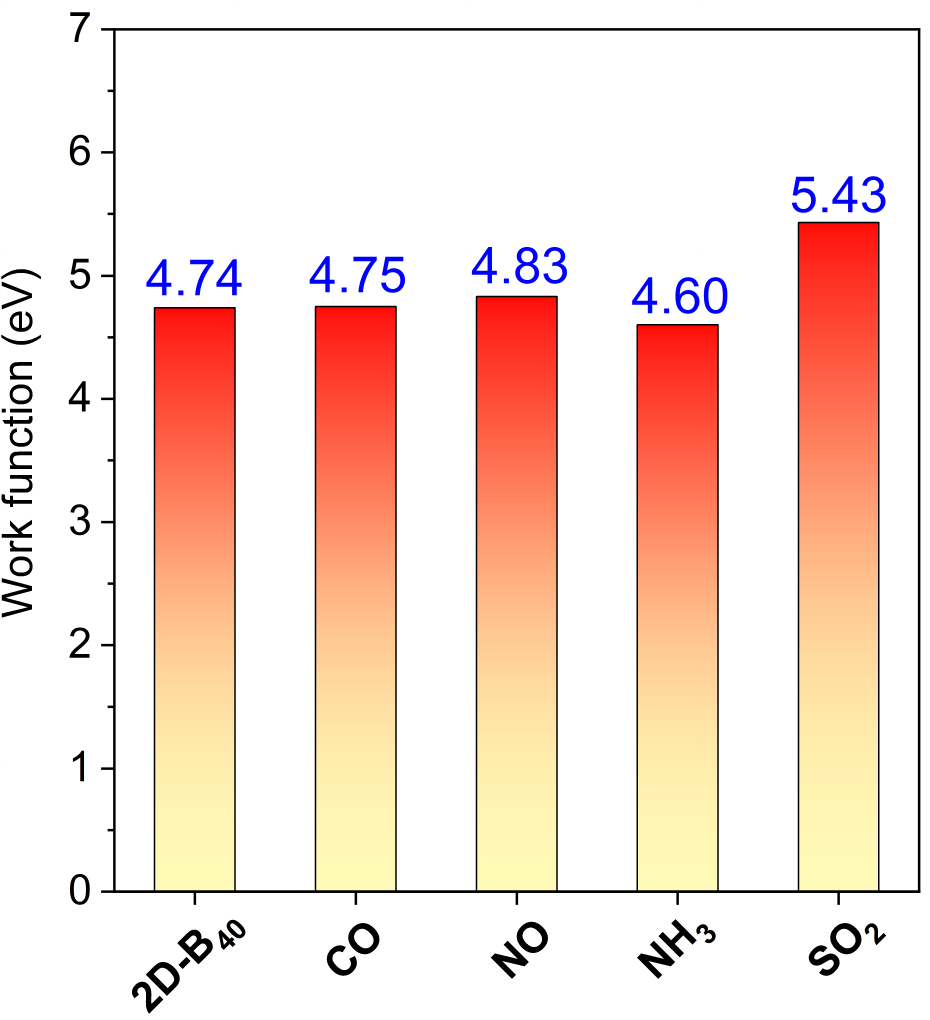}
    \caption{Work function values \(\Phi\) for pristine 2D–B\textsubscript{40} and after adsorption of CO, NO, NH\(_3\), and SO\(_2\)}
    \label{fig:work function}
\end{figure}

\subsection{Gas sensing and capture properties}

Another important descriptor for evaluating gas sensing performance is the work function (\(\Phi\)) of the substrate, which quantifies the minimum energy required to extract an electron from the Fermi level to the vacuum level \cite{martins2025high, elaggoune2025enhancing, rahimi2025b3c2n3, kang2024blue}. In experimental setups, \(\Phi\) is typically measured using Kelvin probe techniques, where the contact potential difference (CPD) reflects variations in surface electronic structure due to adsorption \cite{melitz2011kelvin, nonnenmacher1991kelvin}. Theoretically, the work function is computed as 
\begin{equation}
\Phi = V_{\infty} - E_F,
\end{equation}
\noindent where \( V_{\infty} \) is the electrostatic potential in the vacuum region and \( E_F \) is the Fermi energy of the system.

Figure~\ref{fig:work function} shows the calculated work functions of pristine and gas-adsorbed 2D–B\textsubscript{40}. The pristine surface exhibits a work function of 4.74 eV, consistent with prior reports for borophene under vacuum ($\sim$ 4.9 eV) and on Ag(111) substrates (4.43–4.69 eV) \cite{liu2021nanoscale}. Upon adsorption, distinct shifts in \(\Phi\) are observed. CO induces a negligible increase (4.75 eV), in line with its physisorption character. Conversely, NO and SO\(_2\) adsorption significantly raise the work function to 4.83 eV and 5.43 eV, respectively, due to their electron-accepting behavior and strong charge depletion at the surface (see Figure~\ref{fig:CDD}). In contrast, NH\(_3\), acting as an electron donor, reduces \(\Phi\) to 4.60 eV, consistent with increased surface electron density. These modulations confirm that 2D–B\textsubscript{40} can serve as an electronic sensor platform, where gas-induced shifts in work function translate into detectable electrical responses. 

The relative sensitivity \(S_\Phi\) of the surface can be defined as
\begin{equation}
S_\Phi~(\%) = \left| \frac{\Phi_{\text{ads}} - \Phi_{\text{pristine}}}{\Phi_{\text{pristine}}} \right| \times 100,
\end{equation}
enabling a quantitative comparison of the response to different analytes. The calculated sensitivities are 0.21\%, 1.90\%, 2.95\%, and 14.56\% for CO, NO, NH\(_3\), and SO\(_2\), respectively. These results suggest that 2D–B\textsubscript{40} exhibits promising sensing capabilities, particularly for the chemisorbed species NO, NH\(_3\), and SO\(_2\).

\begin{figure}[htb!]
    \centering
    \includegraphics[width=1\linewidth]{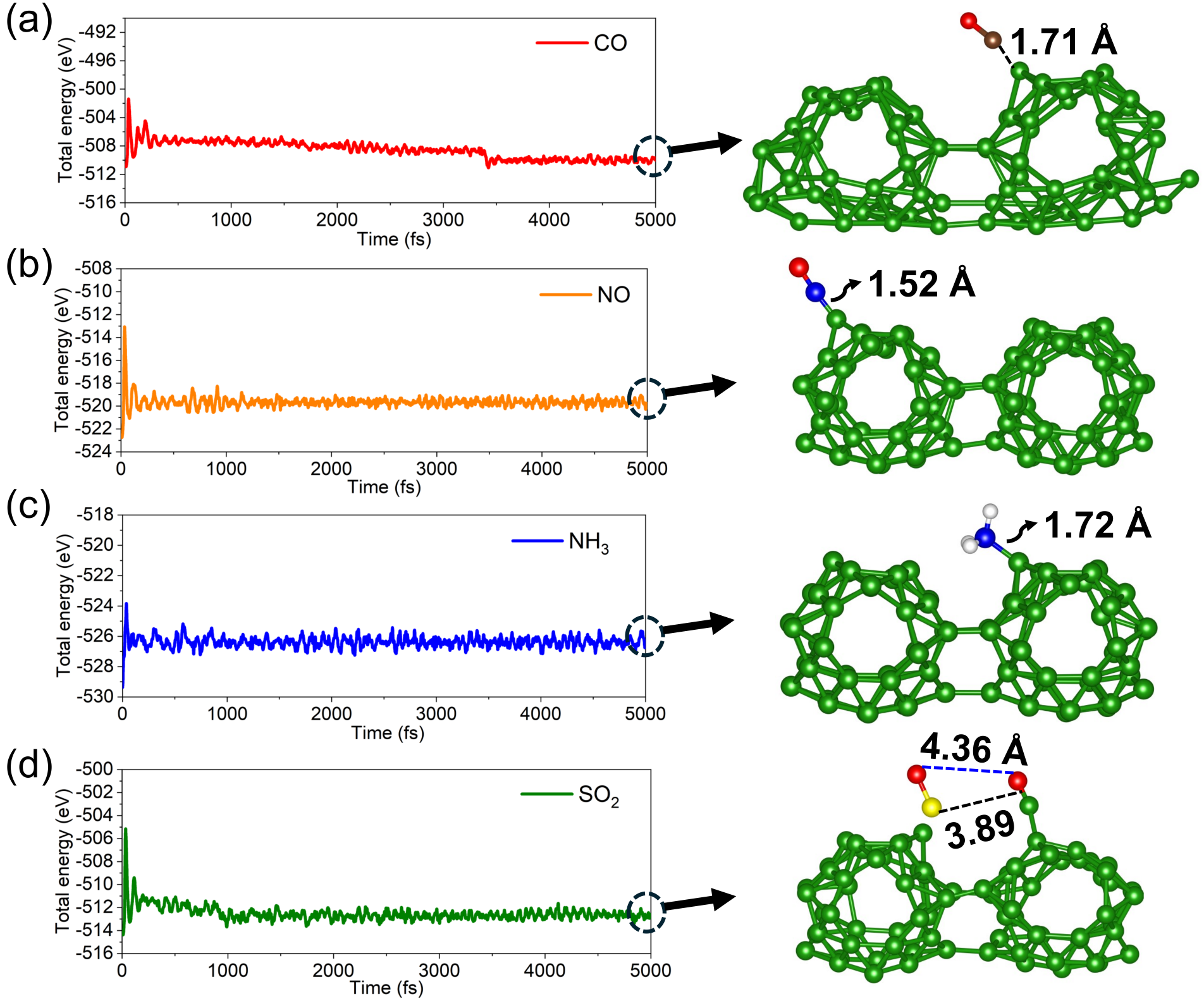}
   \caption{AIMD simulations at 300 K for (a) CO, (b) NO, (c) NH$_3$, and (d) SO$_2$ adsorbed on 2D–B\textsubscript{40}. Left: total energy profiles over 5 ps indicating system stability. Right: final atomic configurations showing adsorption distances.}
  \label{fig:AIMD}
\end{figure}

To assess the thermal robustness and adsorption stability of toxic gas molecules on 2D–B\textsubscript{40}, AIMD simulations were performed at 300 K for 5 ps in the NVT ensemble. Fig.~\ref{fig:AIMD} displays the time-dependent total energy profiles alongside the final atomic configurations of each gas–substrate system. For CO (panel a), although physisorption was initially predicted under static conditions, the molecule forms a stronger B–C interaction (1.71 \r{A}) during AIMD, indicating that thermal energy promotes stabilization and could enhance CO detection at ambient temperature. NO and NH$_3$ (panels b and c) remain stably adsorbed, with only slight elongation of B–N bond lengths to 1.52 and 1.72 \r{A}, respectively, confirming their robust chemisorption on 2D–B\textsubscript{40}.

The simulation of SO$_2$ (panel d) reveals spontaneous molecular decomposition: the S–O bond dissociates during the trajectory, resulting in two B–O distances of 4.36 and 3.89 \r{A}. This suggests that 2D–B\textsubscript{40} can not only sense SO$_2$ but also catalyze its degradation under mild thermal conditions. Such behavior indicates that this material has dual functionality, serving both as a sensing platform and an active surface for environmental remediation.

\section{Conclusions}

Density functional theory simulations were utilized to investigate the structural stability, electronic properties, and gas adsorption performance of a recently proposed two-dimensional boron-based material, 2D–B\textsubscript{40} (borospherene). The results demonstrate that this network exhibits dynamic and thermal stability, a metallic band structure dominated by B($p$) orbitals, and porosity, making it suitable for molecular adsorption.

By examining the interaction of toxic gases, CO, NO, NH$_3$, and SO$_2$, at three distinct adsorption sites, this work established that 2D–B\textsubscript{40} exhibits selective chemisorption of NO, NH$_3$, and SO$_2$, with adsorption energies as strong as $-2.24$ eV and significant charge transfer identified via Bader analysis. CO, in contrast, interacts weakly via physisorption, as evidenced by its minimal adsorption energy and charge transfer. Projected density of states and charge density difference maps further corroborate the orbital hybridization patterns that underpin these distinct binding regimes.

Electronic sensing performance, evaluated through work function analysis, reveals gas-induced shifts of up to 14.6\% for SO$_2$, thereby validating the potential of 2D–B\textsubscript{40} as a highly sensitive detection platform. AIMD simulations confirmed the structural robustness of the system under thermal conditions. Spontaneous dissociation of SO$_2$ was observed, suggesting a dual functional role for 2D–B\textsubscript{40} in both gas detection and catalytic degradation.

\section*{Data Availability}
All data supporting the findings of this study are available within the article.

\section*{Conflicts of interest}
\noindent The authors declare that they have no conflict of interest.

\section*{Acknowledgments}
This work was supported by the Brazilian funding agencies Fundação de Amparo à Pesquisa do Estado de São Paulo (FAPESP) (grants no. 2022/03959-6, 2022/14576-0, 2013/08293-7, 2020/01144-0, 2024/05087-1, and 2022/16509-9), National Council for Scientific, Technological Development (CNPq) (grants no. 307213/2021–8, 350176/2022-1, and 167745/2023-9), FAP-DF (grants no. 00193.00001808/2022-71 and 00193-00001857/2023-95), FAPDF-PRONEM (grant no. 00193.00001247/2021-20), and PDPG-FAPDF-CAPES Centro-Oeste (grant no. 00193-00000867/2024-94). 

\bibliographystyle{unsrtnat}
\bibliography{references}  

\end{document}